\documentstyle[natbib209,aasj,mnfixes,epsfig]{mn-nat}
\begin{document}

\title[Magnetar QPOs]
{On the theory of magnetar QPOs.}

\author[Levin ]{Yuri Levin$^{1,2}$ 
\\
$^1$Leiden University, Leiden Observatory, P.O. Box 9513, NL-2300 RA Leiden
\\
$^2$Leiden University, Lorentz Institute, P.O. Box 9506, NL-2300 RA Leiden}

\date{printed \today}
\maketitle
\begin{abstract}
We consider torsional oscillations of magnetars. 
This problem features  rich dynamics due to the strong
interaction between the normal modes of a magnetar's crust and a continuum
of Magneto-Hydro-Dynamic (MHD) modes in its fluid core.
We study the dynamics
using a simple model of a magnetar possessing a uniform magnetic
field and a thin spherical crust. Firstly, we show that 
global torsional modes only exist when one introduces unphysically
large dissipative terms into the equations of motion; thus global modes are
not helpful for understanding the magnetar Quasi-Periodic Oscillations (QPOs).
%phenomenon.
%Firstly, we assume that neutrons are dynamically coupled to the
%$e-p$-plasma in the core.
%We find that
%due to the presence of MHD continuum in the core, the global normal modes only 
%exist when one 
%introduces sufficiently large friction term $-\gamma d\vec{r}/dt$ 
%into the equations of motion. 
%When the modes do exist, their frequencies are complex, and
%the modes decay over several oscillation periods with the rate independent of 
%$\gamma$. 
%There are 2 important regimes: 
%1. The crustal mode vibrational frequency $\nu$ is
%smaller than those of the MHD continuum. This occurs for crustal
%modes with frequencies $\nu<\sim 60$Hz, if the neutron (super)fluid is dynamically
%decoupled from the charged $e$---$p$ plasma in the core. Then the global modes
%exist which have vibrational frequencies similar (within $\sim 5$\%) to those
%of the purely crustal modes. 
%2. The crustal mode vibrational frequency $\nu$ belongs to those of the
%MHD continuum. 
Secondly, we solve the initial-value problem by simulating the
sudden release of an initially strained crust and
monitoring the subsequent crustal motion. We find that 
the crustal torsional modes quickly
exchange their energy with the MHD continuum in the core, and decay by several
orders of magnitude over the course of $\sim 10$ oscillation periods. After
the initial rapid decay, the crustal motion is stabilized and several time-varying QPOs are 
observed. The dynamical spectrum of the simulated crustal motion is in 
qualitative agreement
with that of the x-ray light-curve in the tail of a giant magnetar flare.
The asymptotic frequencies of some of the QPOs are associated with the special spectral
points---the turning points
or edges---of the MHD continuum, and are not related to those of
the crust. The observed steady low-frequency QPO at 18 Hz 
is almost certainly associated with the lowest frequency of the MHD continuum,
or its first overtone. We also find that drifting QPOs get amplified when they
come near the frequencies of the crustal modes.  This explains why some of the observed QPOs
have frequencies close to the expected crustal frequencies, and why these QPOs are
highly variable with time.

\end{abstract}

\section{introduction}
Magnetar oscillations have recently 
been the subject of  intense theoretical interest.
This interest is motivated by quasi-periodic x-ray luminosity 
oscillations (QPOs), which were
observed in the tails of 2 giant Soft Gamma-ray
Repeaters'
(SGR) flares (Israel et al.~2005, Strohmayer \& Watts 2005, 2006, Watts \& 
Strohmayer 2006, see also Barat et al.~1983). The QPOs typically last for $\sim 100$ seconds,
are detected with 
 high signal-to-noise ratios, and have frequencies which range from $18$ to $1800$Hz. 
The QPOs open an exciting possibility to directly explore the physics of magnetars, and
their correct interpretation is of great importance.

SGR QPOs have been commonly interpreted
as pure crustal shear modes (Duncan 1998, Piro 2005, Lee 2006, Samuelsson \& Andersson 2006,
Watts \& Reddy 2006,
Sotani et al.~2006a). If this interpretation were correct, it would allow one to measure
or strongly constrain the shear modulus and depth of the crust, an unprecedented feat in the
neutron-star astrophysics (Strohmayer \& Watts 2006). However, the presence of the strong magnetic
field which exists inside a magnetar may present difficulties for this interpretation. In particular,   
Levin (2006, L06) has pointed out that hydromagnetic (MHD) mechanical coupling
between the crust and the core occurs on the 
timescale $<0.1$ seconds, and should be taken into account.
L06 made 2 basic points: 1. MHD coupling ensures that pure crustal modes do not exist, and global
modes of the whole star must be considered, and 2. Long-term survival of the global mode is in danger,
since it is expected to couple to a continuum of MHD modes (the Alfven continuum) in the core, and this
coupling would act to damp the mode. More recently, Glampedakis, Samuelsson, \& Andersson 2006 (GSA)
and  Sotani et al.~2006b (S06b) have found global MHD-elastic modes in simple toy magnetar models,
and have argued that the analogues of these modes produce QPOs in real magnetars. 
However, both of the toy models have been constructed in 
a way which explicitly excludes the presence
of an Alfven continuum in the core\footnote{GSA make use of the rectangular geometry
with a uniform magnetic field,
which ensures that all 
of the Alfven modes with the same quantum number have the same frequency. S06b consider dipole
magnetic field in the spherical magnetar, 
but explicitly exclude $l\pm2$ coupling in their equations. This enforces the spherical symmetry in 
the physical problem described 
by their equations of motion.
In fact, their equations are  identical to those that describe oscillation
of the star with a purely radial spherically symmetric magnetic field, and thus, just as in GSA,
the local Alfven waves with the same quantum
numbers have the same frequency. We note that 
Rincon \& Rieutord (2003) and 
Reese, Rincon, \& Rieutord (2004) have previously found a continuum of torsional
modes 
in the MHD configuration 
identical to that in S06b (they have not used any simplifying approximations in their
analysis).
To sum up, in both GSA and S06b the  symmetry of their
toy models collapses the Alfven continuum onto the discrete 
set of frequencies.}.

The coupling of hydrodynamical waves to Alfven continuum has been extensively
studied in the context of solar corona, and is well understood (Ionson 1978, Hollweg 1987a,b). 
The absorption of the
waves by Alfven continuum is sometimes referred to as the {\it resonant absorption}.
In this paper we build on the  work done by the solar physics community and  undertake a thorough investigation 
of the influence of the Alfven continuum on the oscillatory  
behavior of the magnetar crust. The plan of the paper is as follows: 

In the next section we describe our simplified, but topologically correct magnetar
model. We  derive equations of torsional motion, and search for  normal modes of the system.
%First we consider the case when the neutron fluid is tightly coupled to the charged plasma in the
%core, and thus the Alfven continuum begins at the low frequency of $\sim 6$ Hz. In this case the
%frequencies of the discrete crustal modes lie in the range of the continuum.
We  find the normal modes only exist if one introduces
sufficiently large frictional forces, e.g.~the ones of the form
$-\gamma d\vec{r}/dt$, into the equations of motion. 
The eigenfrequencies are complex, and the modes decay with 
the rate independent of $\gamma$, typically over several oscillation periods (i.e., a fraction of a second).
However, the normal-mode analysis is inconclusive, since 
the real frictional forces in a magnetar may be small and the normal modes likely do not exist.
Thus in section 3 we turn to the initial value problem.
We model the continuum by $10^4$ oscillators chosen to mimic closely the MHD dynamics of the
core (one can think of this idea as similar to the one behind spectral codes in fluid dynamics, except that
our equations are linear. 
The true linear behavior of the real magnetar is recovered when the number of oscillators
tends to infinity). We begin the simulations by releasing 
an initially strained crust, and then monitor its motion.
The result of one such simulation is shown in Fig.~9.
We find that the crustal deformation energy is quickly converted 
into the energy of MHD continuum in the core,
as was predicted in L06 and as is suggested by the large imaginary components of the 
normal-mode frequencies obtained in section 2. The amplitude of crustal motion is
reduced by $~10^2$ over several oscillation periods, but is then stabilized as the 
crust reacts to the vigorously-moving
core. {\bf In this second time-interval we find QPOs in the crustal motion}, see Fig.~10. 
The asymptotic frequencies of some of the QPOs are associated with the special spectral points
of the continuum. Both turning points and edges of the continuum produce QPOs 
(see section 3 and Figure 4 for an explanation
of what these are). We also find that when the frequency of a drifting QPO approaches 
that of a crustal mode, the QPOs amplitude gets significantly amplified. Thus crustal frequencies
feature intermittently enhanced power, in agreement with the observations. 
%We find tthat the QPOs  produced by reversals are
%longer lived and generally decay as $t^{-1/2}$, 
%while those from the edges decay as $t^{-1}$, and we provide an analytical
%explanation for this behavior. In some cases we observe 
%icomplicated temporal behavior of the QPOs, presumable due to
%interactions between different branches (overtones) of the Alfven continuum.

%In section 4 we consider the effect of dynamical decoupling of the 
%neutron liquid from a charged plasma, due to 
%potential neutron superfluidity. 
%We find that the net effect is to raise the continuum frequencies by an order of 
%magnitude. This ``saves'' the crustal modes with frequencies less than $\sim 60$Hz. 
%Their frequencies get  modified by
%$\sim 5$\%
%due to  the magnetic-field tension and inertial loading by the core plasma, but remain real and
%the modes don't decay.

In section 4 we present an outlook on the outstanding theoretical questions related to 
magnetar QPOs.

%SMadiscuss 
%the issue of powering the flare itself. We argue  that MHD continuum in the core
%is likely to absorb most of the internal 
%mechanical energy released during the burst, thus creating a potential energy
%budget crisis for producing the flare. Lyutikov (2003) has 
%proposed that instead the flare is powered by a rearrangement
%of the global twisted magnetosphere configuration. In Lyutikov's scenario
%the energy is released directly into the magnetosphere
%via the field-lines reconnection, thus avoiding the energy budget problem we describe.

%We conclude and present our outlook in section 6. 

\section{Basic model and its normal modes.}
Finding oscillatory modes of magnetic stars presents a formidable computational
and conceptual problem. Rincon \& Rieutord (2003) and 
Reese, Rincon, \& Rieutord (2004) in a tour-de-force
calculation have computed normal modes of an incompressible fluid shell threaded
with a dipole magnetic field. Partly motivated by their work, we choose a simple
magnetar model, with several basic assumptions: 
1. We take the elastic crust to be thin compared to the core size.
This is an excellent approximation for mode frequencies less than $\sim 600\hbox{Hz}$.
2. We assume that the fluid core has uniform density and is threaded by 
a homogeneous internal magnetic field directed along the $z$-axis.
This assumption does not change the physics of the problem, but does simplify the calculations and
makes them more transparent.

We also consider oscillations with purely asimuthal, $\phi$-independent displacements $\xi$ and
$\bar{\xi}$ of the core and  the crust, respectively:
\begin{eqnarray}
\xi_r&=&\xi_\theta=\bar{\xi}_r=\bar{\xi}_{\theta}=0,\nonumber\\
\xi_\phi&=&\xi_\phi(r,\theta),\label{xi1}\\
\bar{\xi}_\phi&=&\bar{\xi}_{\phi}(\theta)\nonumber
\end{eqnarray}
Here $r,\theta,\phi$ are the spherical coordinates.  We have made use of the thin-crust assumption
in writing the last equation.

%We shall now introduce the equations of motion in 3 stages. First, we will consider the motion
%of the crust in isolation from the core, and rederive crustal normal-mode fr
%We now write down the equation of motion for the crustal displacement:
We are now in the position to write down  equations of motion for small displacements
of the crust and the core:
\begin{eqnarray}
{\partial^2\bar{\xi}_\phi\over \partial t^2}&=&L_{\rm el}(\bar{\xi}_\phi)+L_B,\label{barxi}\\
{\partial^2\xi_\phi\over \partial t^2}&=&c_a^2
\left({\partial^2\xi_\phi\over\partial z^2}\right)_{r\sin\theta}-\gamma{\partial \xi_\phi
\over \partial t}.
\label{xi2}
\end{eqnarray}
where the partial derivative on the right-hand side of Eq.~(\ref{xi2}) is evaluated along the
$z$-direction. Here $c_a$ is the Alfven velocity, $L_{\rm el}(\bar{\xi}_\phi)$ 
and $L_B$ is the acceleration of the crust due to the elastic and magnetic stresses, respectively.  
We have introduced the frictional term $-\gamma \partial \xi_\phi/\partial t$ into 
Eq.~(\ref{xi2}); we will show shortly that this term is needed to regularize the resonant response
of the core to the periodic motion of the crust and is crucial for the existense of a normal mode.
The expression for $L_B$ is derived below, while that for 
$L_{\rm el}$ is derived in the Appendix:
\begin{equation}
L_{\rm el}(\bar{\xi}_\phi)=\omega_{\rm el}^2\left[{\partial^2\bar{\xi}_\phi\over\partial \theta^2}+
\cot(\theta){\partial\bar{\xi}_\phi\over\partial\theta}-(\cot^2(\theta)-1)\bar{\xi}_\phi
\right], \label{L}
\end{equation}
where $\omega_{\rm el}$ is the frequency given by 
\begin{equation}
\omega_{\rm el}={\sqrt{\bar{\mu}/\bar{\rho}}\over R},
\label{omegael}
\end{equation}
where $\bar{\mu}$ and $\bar{\rho}$ are the vertically averaged shear modulus and density of
the crust, respectively, and $R$ is the radius of the star.

\subsection{Dynamics of the core: continuum of modes and response to periodic forcing.}
It is instructive to consider the motion of the core fluid, with the assumption of a fixed or 
periodically moving crust as an external boundary condition. The former will elucidate
the structure of the Alfven continuum, while the latter is instrumental in the normal-mode
analysis.

Hydro-magnetic stress enforces a no-slip boundary condition at the crust-core interface:
\begin{equation}
\xi_{\phi}(R,\theta)=\bar{\xi}_\phi.\label{boundary1}
\end{equation}
Let us consider the dynamics of the core, with the assumption of a fixed crust and zero friction.
The core displays a continuum of singular oscillatory
solutions to the pair of equations (\ref{xi2}) and (\ref{boundary1}). These solutions are
localized on cylinders of radius $\eta_0$, with $0<\eta_0<R$, and their mathematical form is
expressed most easily in cylindrical coordinates $\eta, z, \phi$:
\begin{eqnarray}
\xi_{\phi}(\eta,z,\phi,t)&=&\delta(\eta-\eta_0)\sin[n\pi z/h(\eta_0)]\times\nonumber\\
                         & &\exp[i\pi n c_a t/h(\eta_0)]
\label{oddcore}
\end{eqnarray}
or
\begin{eqnarray} 
\xi_{\phi}(\eta,z,\phi,t)&=&\delta(\eta-\eta_0)\cos[(n+1/2)\pi z/h(\eta_0)]\times\nonumber\\
                         & &\exp[i\pi n c_a t/h(\eta_0)];
\label{evencore} 
\end{eqnarray}
cf.~section 3 of L06.
Here $h(\eta_0)=\sqrt{R^2-\eta_0^2}$ is the height of each cylinder, and $n$ is an integer.
While this continuum of MHD modes was derived for a simple magnetic-field geometry, it must
exist in other field geometries which can be obtained by continuous deformation of the uniform
field. Moreover, since a continuos deformation
of the field  changes the mode frequencies continuously, the topology of the spectrum remains
unchanged. This means that the modes will form a countable set of one-dimensional continua;
in other words, even for a complicated 
magnetic-field configuration 
the mode is parametrized by a pair of  real and integer numbers. The latter
consideration will become important when we discuss general properties of QPOs in section 3.

Now consider the core's response to a periodic motion of the crust, 
$\bar{\xi}(\theta)=g(\theta)\exp(i\omega t)$, where $\omega$ could, in general, be complex.
First we note that the geometry of our problem possesses the reflection symmetry with respect
to the $z=0$ plane. Therefore the normal modes will be either even or odd with respect to $z$,
and we restrict the crustal motion to that with $g(\theta)=\pm g(\pi-\theta)$.
From Eqs.~(\ref{xi2}) and (\ref{boundary1}), we see that in the ``odd'' case
the core motion is given in cyllindrical coordinates by
\begin{equation}
\xi_\phi(\eta,z)={\sin(kz)\over \sin[kh(\eta)]}\bar{\xi}_\phi[\theta(\eta)],
\label{xi3}
\end{equation}
while in
the ``even'' case
\begin{equation} 
\xi_\phi(\eta,z)={\cos(kz)\over \cos[kh(\eta)]}\bar{\xi}_\phi[\theta(\eta)],
\label{xi4}
\end{equation}
where $\theta(\eta)=\arcsin(\eta/R)$, $h(\eta)=\sqrt{R^2-\eta^2}$, and $k=\sqrt{\omega^2-
i\omega\gamma}/c_a$.

\subsection{Normal modes}
Now we are ready to derive the acceleration of the crust due to the hydromagnetic stress
at the crust-core interface:
\begin{equation}
L_B=-{\rho_c c_a^2\over \Sigma}\cos\theta \left({\partial\xi\over \partial z}\right)_{z=h},
\label{ab1}
\end{equation}
where $\rho_c$ is the density of the fraction of the core fluid which participates
in the Alfven motion, $\Sigma$ is the column density of the crust, and
the partial derivative is evaluated at $(\eta,z)=(R\sin\theta, R\cos\theta)$.
By substituting Eq.~(\ref{xi3}) or (\ref{xi4}) into Eq.~(\ref{ab1}), we get
the following expressions for magnetically-driven acceleration of the crust:
\begin{equation}
L_B(\theta)=-\nu_a\omega_1 {\rho_c R\over \Sigma} \cot(\omega_1\cos\theta/\nu_a)
            \cos\theta \bar{\xi}_\phi(\theta)
\label{ab4}
\end{equation}
for the odd modes,
and
\begin{equation}
L_B(\theta)=\nu_a\omega_1 {\rho_c R\over \Sigma} \tan(\omega_1\cos\theta/\nu_a)
            \cos\theta\bar{\xi}_\phi(\theta)
	    \label{ab5}
\end{equation}
for the even modes. Here $\nu_a=c_a/R$, and $\omega_1=\sqrt{\omega^2-i\omega\gamma}$.

We can now see that the  Eq.~(\ref{barxi}), together with Eqs.~(\ref{L}) and (\ref{ab4}/\ref{ab5}),
form an ordinary second-order differential equation for $\bar{\xi}_\phi(\theta)$. The values of
$\omega$ get selected by requiring that the solution of  Eq.~(\ref{barxi}) satisfies the boundary
conditions\footnote{These boundary conditions are derived mathematically
from the Frobenius series expansion of $\bar{\xi}_\phi(\theta)$ near the poles,
or physically by requiring the crustal angular velocity and acceleration be finite at the poles.}
at the poles $\theta=0,\pi$:
\begin{equation}
\bar{\xi}_\phi(\theta)={\partial^2\bar{\xi}_\phi(\theta)\over\partial\theta^2}=0.
\label{boundary3}
\end{equation}
We found it practical to solve Eq.~(\ref{barxi}) in the upper hemisphere, but require
that at the equator either $\bar{\xi}_\phi(\theta)=0$ for the odd modes,
or $\partial\bar{\xi}_\phi(\theta)/\partial\theta=0$ for the even modes.
We also found it useful to make the substitution 
\begin{equation}
q(\theta)=\bar{\xi}_\phi(\theta)/\theta,
\end{equation}
and rewrite the equations in terms of $q(\theta)$. The new equations do not have
a singularity at the pole $\theta=0$, and are very easy to integrate on the
computer. We have checked the code by solving both analytically and numerically
the case with $L_B=0$. We find analytically that 
the wavefunctions of the free-crust vibrational modes are given by
\begin{equation}
\bar{\xi}_\phi(\theta)=\partial Y_{l0}(\theta)/\partial \theta,
\label{freecrust}
\end{equation}
with the eigenfrequencies 
\begin{equation}
\omega_l=\sqrt{(l-1)(l+2)}\omega_{\rm el}.
\label{omfree}
\end{equation}
This $l$--scaling is in full agreement with that of Samuelsson \& Andersson (2006). Our numerics
gives excellent agreement with these results.

Before we discuss our numerical results for the case of a non-zero $L_B$, one qualitative
remark is due. From Eqs.~(\ref{ab4}) and (\ref{ab5}), we see that if $\omega$ is
real and the friction coefficient 
$\gamma=0$, then $L_B$ diverges for the values of $\theta$ which
correspond to the location of the Alfven continuum mode in resonance with $\omega$.
It is these resonant interactions that are largely responsible for the exchange of energy
between the global vibration and the Alfven cantinuum, and that are thus responsible for
determining the imaginary part of $\omega$ (hence the name
``resonant absorbtion'' in the solar literature). In our Runge-Kutta routine, we enforce
small $\theta$-steps which scale as $L_B^{-2}$ near the singularities (and we  check that
our results do not change when the step-size is reduced by a factor of 10).

We now discuss our  numerical results for normal modes of a magnetar. 
%In this section, we shall
%concentrate on the case where the core fluid moves as a single fluid, and will postpone the discussion
%of the alternatives (e.g., the case of fully decoupled neutron superfluid) until section 4.
%When all of the fluid core is tightly coupled to the field lines,
%the Alfven continuum begins at $\sim 6$Hz, and all of the
%crustal frequencies lie inside the range of the continuum. 
%This is the regime in which resonant absorption occurs. i
We find that the
normal modes exist only when $\gamma$ is sufficiently large, i.e.~$\gamma>\gamma_{\rm crit}$.
When this condition is satisfied, then the frequency of the normal mode is complex and
the mode decays with the rate close to $\gamma_{\rm crit}$. By contrast, when $\gamma<\gamma_{\rm crit}$,
a thorough numerical search fails to identify the complex eigenfrequency for which all of
the boundary conditions are satisfied.
This is in full agreement with  previous work done on the resonant absorption in the solar
corona. For example, in Steinolfson's (1985) simulation the system behaves like a decaying
normal mode when the friction is sufficiently large, while for small friction no normal-mode-like
behaviour occurs and instead, the phase mixing is observed,
where individual modes of the continuum are excited and oscillate each at their
own frequency. The same occurs in our initial-value simulations,
which we describe in Section 3. Hollweg (1987a,b) gives an excellent physical
discussion of why, when the mode exists, its decay-rate is friction-independent\footnote{Briefly, this
can be understood as follows: the energy is absobed in narrow resonant layers; in our case they have 
cylindrical shapes. Friction produces 2 main effects: 
(a) it reduces the excitation amplitude of the resonant layer,
and (b) it increases the effective width of the layer. 
It is easy to check that for the simple frictional terms we are using, the 2 effects compensate each
other and the total absoption power is friction-independent. Hollweg (1987) proves that the same result
holds for more complicated dissipative effects, like viscosity or Ohmic dissipation in the plasma.}

In figure 1 we show the decay rate for the lowest-frequency normal mode as a function of $\gamma$,
for the ten values
magnetic-field strengths $B_{\rm eff}=\sqrt{4\pi\rho} c_a$ ranging from $10^{14}$ tp $10^{15}$
Gauss. 
\begin{figure}
\begin{center}
%\hbox{
\epsfig{file=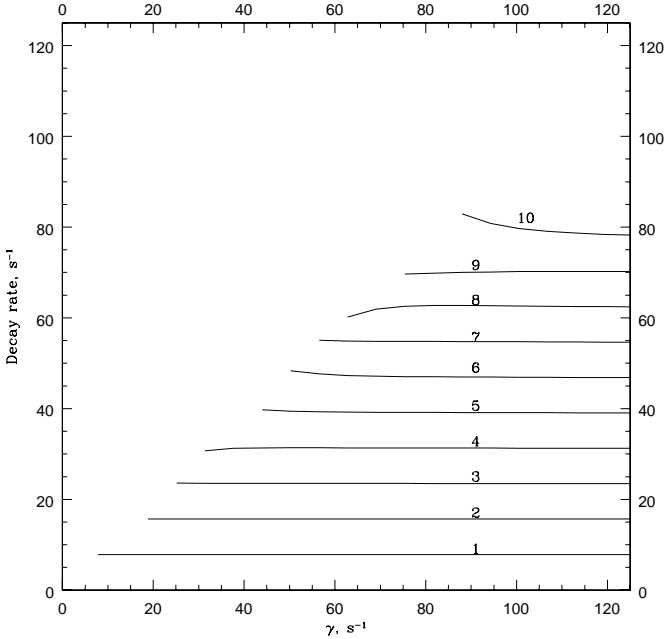, width=9cm}
%}
\end{center}
\caption{Damping rate for a global torsional mode is plotted as a function of the dissipative rate
$\gamma$. The lines in the figure are numbered by an integer $i=1,...,10$; each lline corresponds to the 
magnetic field strength $B_{\rm eff}=i\times 10^{14}$G. For small $\gamma$, i.e.~to the left of the lines, 
the modes don't exist.}
\end{figure}

Here $B_{\rm eff}=\sqrt{BB_c}$ if the protons form a superconductor
with the critical field strength $B_c$, and $B_{\rm eff}$ if the protons form a normal fermi liquid, and
$\rho$ is the density of the core material coupled to the magnetic field\footnote{This density can range from
the density of protons $\rho_{\rm pr}$ if the neutron superfluid is entirely decoupled from the proton motion,
to the full core density $\rho_{\rm core}$ if the neutrons are efficiently coupled to protons. Our guess is that in
 reality $\rho$ takes the value somewhere in between the 2 extremes: it may be hard for neutron superfluid
 to become completely decoupled from the charged component, since neutron superfluid vortices are expected 
 to be strongly magnetized and may interact strongly with the superconducting flux-tubes; see, e.g., Ruderman
 et al.~1998 and references therein. We take the numerical value $\rho=10^{14}\hbox{g}/\hbox{cm}^3$, about
 $1/10$ of the core density and twice the proton density. Our choices for $B$ and $\rho$ affect directly the 
 value of the lowest-frequency QPO simulated in section 3.}.  
Our fiducial parameters were $\rho R/\Sigma=10$ [probably an underestimate,
but its increase would only increase the crust-core coupling---see Eqs (\ref{ab4})
and (\ref{ab5})], $\omega_{\rm el}=2\pi\times 20$Hz,
and $\nu_a/\omega_{\rm el}=0.2$, appropriate for $\rho\sim 10^{14}\hbox{g/cm}^3$. We see that when the
modes exist they show rapid decay, on the timescale $\ll 1$ second. This result holds
for all higher-order modes we have considered, and shows 
how efficiently the crustal motion is coupled to the continuum. However, the friction (e.g., due
to viscosity)
may be small in magnetars, and the normal modes most likely do not exist. Thus the problem of
magnetar torsional motion must be addressed using initial-value simulations. This is the subject of
the next section.

\section{Initial-value simulations}
The aim of this section is to simulate torsional motion of a magnetar. During this motion, the
discrete torsional modes of the crust interact strongly with a continuum of Alfven modes in the
core, and this interaction affects dramatically the motion of the crust. In the next subsection,
we explore with a help of a toy model the dynamics of a harmonic oscillator coupled to
to a continuum of oscillators. The toy model provides us with an insight into QPOs of such a system,
and gives us intuition for what to expect in the case of a magnetar. In subsection 3.2 we
set up the initial-value simulation for our magnetar model (the ``real'' magnetar, as 
opposed to the toy model in subsection 3.1), and present results.

\begin{figure}
\begin{center}
%\hbox{
\epsfig{file=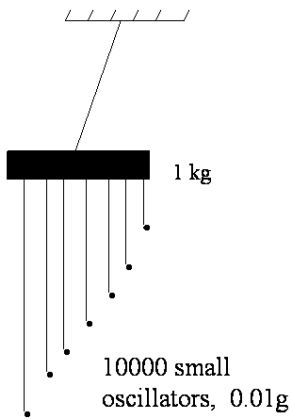, width=9cm}
%}
\end{center}
\caption{Big pendulum coupled to a large number of small one.}
\end{figure}

\subsection{Coupling of a harmonic oscillator to a continuum: toy model.}
In Figure 2 we show the set-up of our toy problem. We consider a pendulum weighing 1kg, with
a proper oscillation angular
frequency of $\omega_0=1$ (the units are irrelevant). We model the pendulum
coupling to a continuum of modes by suspending 10000 tiny pendulae, each weighing $0.01$g,
from the big pendulum. We arrange the angular frequency of the small pendulae to be
\begin{equation}
\omega_m=0.5+.0001*m,
\label{dispersion1}
\end{equation}
where $m=1,2,...,10000$.
Thus the frequency of the big pendulum lies in the middle of the range of those of the small
pendulae. The initial condition for our 
simulation is as follows: the big pendulum is deflected by a small angle (we
keep the problem linear), while the small pendulae are relaxed and hanging straight down.
This is meant to mock an initially strained crust and relaxed core. The big pendulum
is  released, and the evolution is followed by two independent numerical
routines. One routine uses the 4-th order Runge-Kutta method, while the other one
uses a symplectic second-order leapfrog algorithm, which is very robust for simulating
Hamiltonian systems (see, e.g., Kinoshita, Yoshida, \& Nagai 1991). 
Both runs conserve the total energy of the system with high precision, and produce results which are
in excellent agreement with each other.
\begin{figure}
\begin{center}
%\hbox{
\epsfig{file=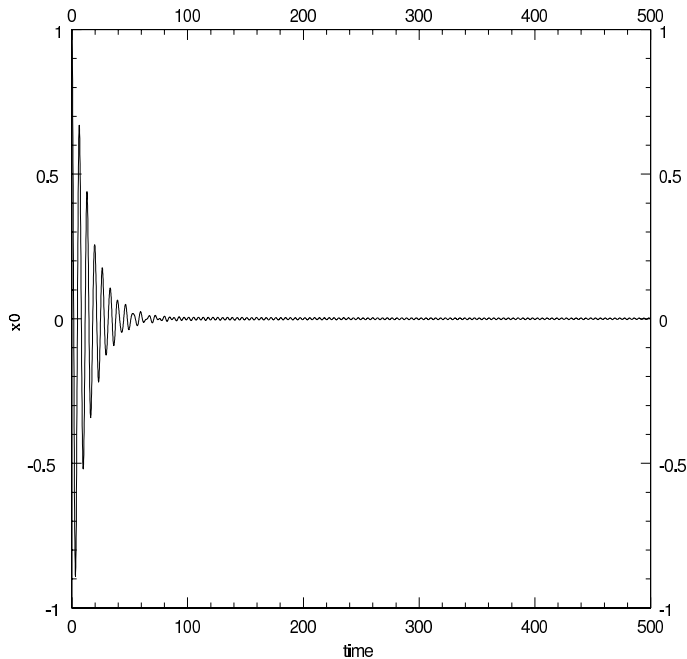, width=9cm}
%}
\end{center}
\caption{Big-pendulum displacement as a function of time.}
\end{figure}
In Fig.~3 we plot the big-pendulum displacement as a function of time. After several oscillations,
the amplitude of the pendulum's motion is reduced by $\sim 100$, as the energy is rapidly transfered
from the big pendulum to the small ones. Then the exponential decay abruptly stops,
as the big pendulum now reacts to the collective  pull of the small ones. The blow-up of this
second region is shown in Fig.~4. The amplitude still decays, but only slowly, as $1/t$.
\begin{figure}
\begin{center}
%\hbox{
\epsfig{file=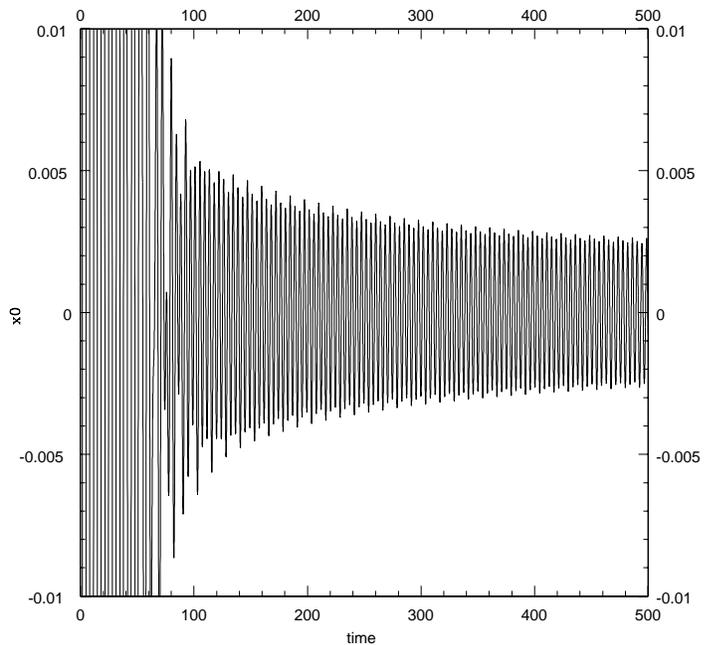, width=9cm}
%}
\end{center}
\caption{Zoom-in on the post-decay part of Fig.~3. Quasi-periodicity is clearly visible.}
\end{figure}
Astonishingly, even with naked eye, one can detect {\bf QPO(s)} in the pendulum motion!
Figure 5 shows the time Fourier transform of the big-pendulum displacement for this
interval of time. 
\begin{figure}
\begin{center}
%\hbox{
\epsfig{file=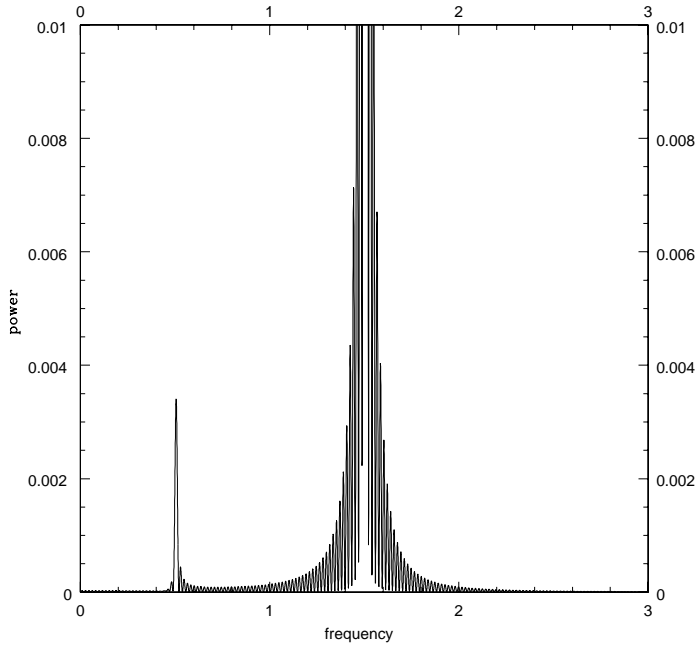, width=9cm}
%}
\end{center}
\caption{Power spectrum of the post-decay big-pendulum displacement. Two
QPOs are clearly visible; they are associated with the edges of the continuum
frequency interval.}
\end{figure}
Two QPO frequencies are clearly present, $1.5$ and $0.5$, both
identified with the edges of the continuum and NOT with the natural frequency of the big pendulum!

We get a clue for the origin of these QPOs by plotting the phases of small pendulae as a function
of the oscillator  frequency, $\omega_m$; 
see Figure 6. Over the range of $m$, the phases average out, thus preventing the small
pendulae from
pulling coherently on the big pendulum. 
\begin{figure}
\begin{center}
%\hbox{
\epsfig{file=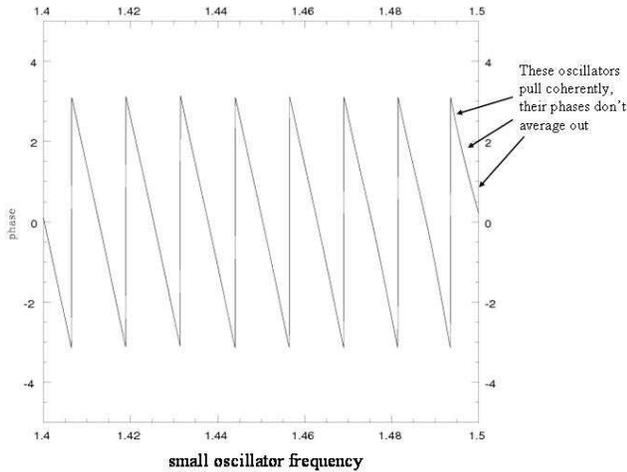, width=9cm}
%}
\end{center}
\caption{Phases of the small pendulae.}
\end{figure}
The only location where this averaging does not occur
is near the end points of the spectrum; see the arrows on Figure 6. Thus the pendulae near
the end points of the spectrum do pull coherently on the big pendulum and produce the 2
QPOs observed in the simulations. The number of ``coherent'' oscillators shrinks as $1/t$,
as the phase gradients with respect to $m$ grow linearly with time. This explains why the QPO
amplitude decays as $1/t$. In Appendix B we present a more mathematical way to understand this 
phenomenon.

Apart from edges, there may be other special points in the continuum which could generate QPOs.
One example is the local maximum or minimum in $\omega_m$ as a function of $m$; we shall
call such special points {\it the turning points}. The same reasoning as that for the edges,  shows
that the phases of small oscillators near the turning point will not average out for some time, and
hence these oscillators will act coherently. Moreover, the density of states is higher
near a turning point than that near an edge, and diverges as $|\omega-\omega_0|^{-1/2}$, where
$\omega_0$ is the frequency of the reversal. We thus expect that turning points generate stronger
QPOs than the edges; this is confirmed by our simulations shown below. The simulations also show
that the turning-point-generated QPOs are longer lived than the edge-generated, and their amplitudes
decay only as
$t^{-1/2}$. This decay law is explained mathematically in Appendix B. Figure 7 shows an example
of a spectral law with a turning point. 
\begin{figure}
\begin{center}
%\hbox{
\epsfig{file=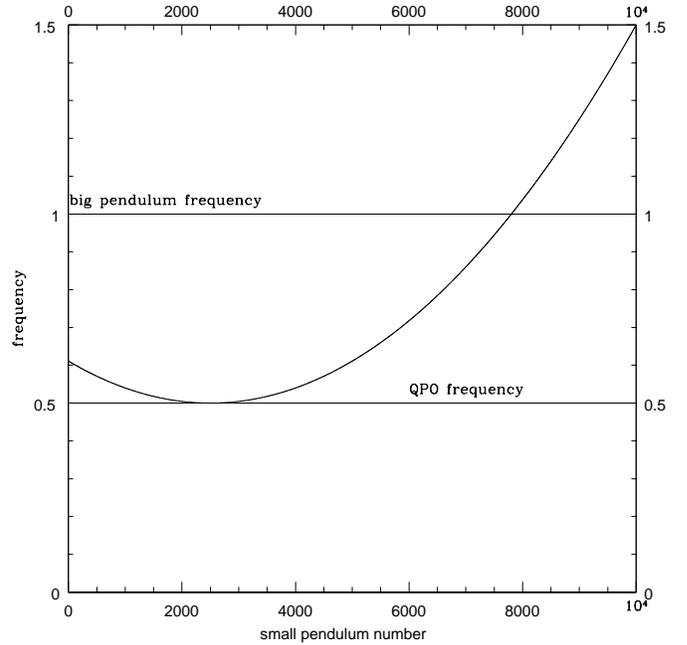, width=9cm}
%}
\end{center}
\caption{Spectral law with the turning point.}
\end{figure}

We simulate the initial-value problem for this example and show
in Fig.~8 the big pendulum's displacement as a function of time. The strong 
long-lived QPO at the turning-point frequency of $0.5$ is apparent with
the naked eye.
\begin{figure}
\begin{center}
%\hbox{
\epsfig{file=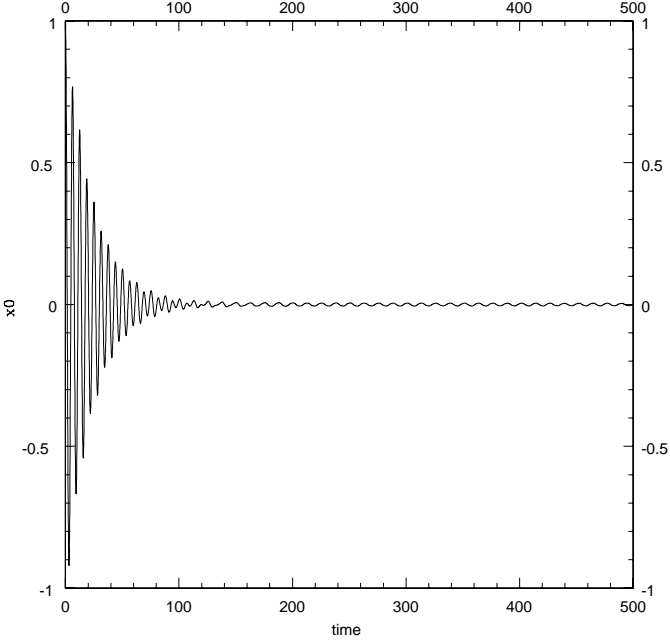, width=9cm}
%}
\end{center}
\caption{Big-pendulum displacement as a function of time. Here it is calculated
for the case when the small pendulae follow the spectral law in Fig.~7. QPO associated with the turning
point is clearly visible in the post-decay time interval.}
\end{figure}

\paragraph{Effect of viscosity.} We have modeled the effect of viscosity by introducing
frictional forces between neighbouring small oscillators. We expect that with the passage of time
the small oscillators get more out of phase, the velocity shear increases and so does the dissipation
rate. This is confirmed by our simulations: the total mechanical energy of the system is drained
efficiently after some time. In magnetars, this may efficiently heat the core and affect the
post-burst afterglow; we shall discuss these issues elsewhere. We find, though, that the QPOs
survive for a much longer time than the global mechanical energy, because the oscillators creating it 
are precisely those ones which are moving in concert. The QPOs generated by the turning points are
particularly robust.

\subsection{Initial-value problem for the magnetar model}
Lessons learned from the toy model lead us to expect (a) rapid decay of initial
crustal perturbation and excitation of the core continuum, and (b) QPOs generated by the edges and 
turning points of this continuum. In our model for the magnetar we have turning points in all
Alfven overtones, at angular frequencies $\omega_n=n\pi c_a/R$ for the odd modes, and
$\omega_n=(n-1/2)\pi c_a/R$ for the even ones; the odd torsional motions are decoupled
from the even ones. Thus potentially we expect QPOs at all of these frequencies. We shall find however,
that our magnetar model displays a much richer dynamics than the toy models of the previous subsection,
although the basic features of the toy models (initial rapid decay of the crustal motion,
QPOs associated with continuum turning points) remain. In what follows
we explain how our initial-value simulations are set up and show the results.
\paragraph{Modal decomposition.}
A crustal displacement could be represented as a sum of the crustal normal modes:
\begin{equation}
\bar{\xi}_{\phi}(\theta, t)=\Sigma_j b_j(t) f_j(\theta),
\label{nm1}
\end{equation}
where $f_j(\theta)$ are proportional to the functions given in Eq.~(\ref{freecrust}),
and are normalized so that
\begin{equation}
\int_0^\pi f_i(\theta)f_j(\theta) d\theta=\delta_{ij}.
\label{normalization}
\end{equation}
We now develop a formalism which allows us to numerically compute the time-evolution of $b_i(t)$.
When the crust is not coupled to the core, $b_i(t)$ oscillate harmonically with
the fruequencies of corresponding crustal modes, but their behaviour is very different when 
the crustal modes are coupled to the continuum modes in the core. In what follows we 
model this continuum with a large but finite number of small oscillators, and we test that our
results do not change when the number of oscillators is increased.

Consider, for concreteness, only odd modes (they remain decoupled from the even ones because
of the reflection symmetry). Recall that for a fixed cyllindrical radius $\eta_0$, the
equation of motion for the core fluid is
\begin{equation}
{\partial^2\xi_{\phi}(\eta_0,z)\over\partial t^2}=c_a^2{\partial^2\xi_{\phi}(\eta_0,z)\over
\partial z^2},
\label{alf}
\end{equation}
with  boundary conditions $\xi_{\phi}(\eta_0,h)=\bar{\xi}_{\phi}[\theta(\eta_0)]$
and $\xi_{\phi}(\eta_0,0)=0$,
where $h=\sqrt{R^2-\eta_0^2}$ is the cyllinder's height, and $\theta(\eta_0)=
\arcsin (\eta_0/R)$. Let us introduce a new variable,
\begin{equation}
\chi(\eta_0,z)=\xi_\phi(\eta_0,z)-\bar{\xi}_\phi[\theta(\eta_0)]{z\over h}.
\label{chi}
\end{equation}
We obtain an inhomogeneous equation for $\chi$,
\begin{equation}
{\partial^2\chi(\eta_0,z)\over\partial t^2}-c_a^2{\partial^2\chi(\eta_0,z)\over
\partial z^2}=-{z\over h} {\partial^2\bar{\xi}_\phi[\theta(\eta_0)\over
\partial t^2},
\label{inhom1}
\end{equation}
but with easy boundary conditions
\begin{equation}
\chi(\eta_0,0)=\chi(\eta_0,h)=0.
\label{bound1}
\end{equation}
Because of this boundary conditions, we can expand $\chi$ in a Fourier series:
\begin{equation}
\chi(\eta_0,z,t)=\Sigma_{n=1}^{\infty} a_n(\eta_0,t)\sin(n\pi\bar{z}),
\label{chi1}
\end{equation}
where $\bar{z}=z/h(\eta_0)$. The right-hand side of Eq.~(\ref{inhom1}) can be expanded
in the same Fourier basis using the identity
\begin{equation}
\bar{z}=-2\Sigma_{n=1}^{\infty}{(-1)^n\over n\pi}\sin(n\pi\bar{z}).
\label{fourid}
\end{equation}
Now, by substituting Eqs.~(\ref{chi1}) and (\ref{fourid}) into
Eq.~(\ref{inhom1}), we obtain the equations of motion for $a_n$:
\begin{equation}
{\partial^2 a_n(\eta_0,t)\over \partial t^2}+n^2\pi^2\nu_a^2
 a_n(\eta_0,t)={2(-1)^n\over \pi n}{\partial^2\bar{\xi}_{\phi}[\theta(\eta_0)]
 \over\partial t^2}.
 \label{eqmot1}
 \end{equation}
We now use Eq.~(\ref{ab1}) to write down the expression for the hydromagnetic backreaction 
on the crust:
\begin{equation}
L_B(\theta)=-{\rho R\over \Sigma}(c_a/R)^2\left(\Sigma_{n=1}^{\infty} n\pi(-1)^n 
a_n[\eta(\theta)]+\bar{\xi}(\theta)\right).
\label{lbmodel}
\end{equation}
The crustal mode amplitudes $b_m(t)$ obey the following equations of motion:
\begin{equation}
{\partial^2 b_m\over \partial t^2}+\omega_m^2 b_m=\int_0^\pi L_B(\theta)f_m(\theta)
\sin\theta d\theta,
\label{mainmodal1}
\end{equation}
where $\omega_m$ is the frequency of the $m$'s crustal mode. In writing the last
equation, we have used the normalization property of the modal wavefunctions 
$f_m(\theta)$. So far we have not used any approximations. Now we discetize the
intergral in Eq.~(\ref{mainmodal1}) by summing over a large number $N$ of points
$\theta_i$, which we take to be equally spaced with the interval $\Delta\theta=\pi/N$.
When we do the sum, the continuum mode aplitude $a_n(\theta)$ is substituted
with the discrete one, $a_{in}=a_n(\theta_i)$. Thus effectively doing the sum
instead of the integral substitutes a continuum of the core modes with the large
number of the discrete core modes. The true continuum dynamics is fully recovered
when $N$ goes to infinity.

We are now in the position to write down the equations of motion for the coupled
crustal and core modes. From Eq.~(\ref{eqmot1}), we have
\begin{equation}
{\partial^2 a_{in}\over \partial t^2}+n^2\pi^2\nu_a^2
 a_{in}={2(-1)^n\over \pi n}\Sigma_{m=1}^{\infty}{\partial^2 b_m(t)\over 
 \partial t^2} f_m(\theta_i).
 \label{firsteq}
 \end{equation}
 Further, from Eq.~(\ref{mainmodal1}) we get
\begin{eqnarray}
{\partial^2 b_m\over \partial t^2}&+&
\left(\omega_m^2+{\rho R\over \Sigma}\nu_a^2\right) b_m=\nonumber\\
&=&
-{\rho R\over \Sigma}\nu_a^2\Delta\theta \Sigma_{n,i} n\pi(-1)^n f_m(\theta_i)
\sin\theta_i a_{in}.
\label{secondeq}
\end{eqnarray}
Equations (\ref{firsteq}) and (\ref{secondeq}) are the main equations of this
section. As with our toy models, we integrate these equations using 2 independent numerical
techniques, the fourth-order Runge-Kutta and the second-order leapfrog. The 2 methods give results
which are in excellent agreement with each other.

We can now show some results from our numerical experiments. Here, we consider the 2 lowest
odd crustal modes, $b_1$ and $b_2$, with frequencies of $40$Hz and $84.5$Hz, coupled to 10000 odd modes
of the continuum: 1000 values of $\theta_i$ with 10 Alfven overtones at each point.
This model should be representative for variability
below $100$Hz. For higher frequencies it is desirable to move away from the thin-crust approximation;
this is the subject of future investigations. We begin with the 
crustal displacement $b_1=b_2=1$ and the relaxed core, $a_{in}=0$. We monitor the crustal displacement
at $\theta=\pi/4$. In Fig.~9 we plot this displacement as a function of time for the first few seconds after
release. 
\begin{figure}
\begin{center}
%\hbox{
\epsfig{file=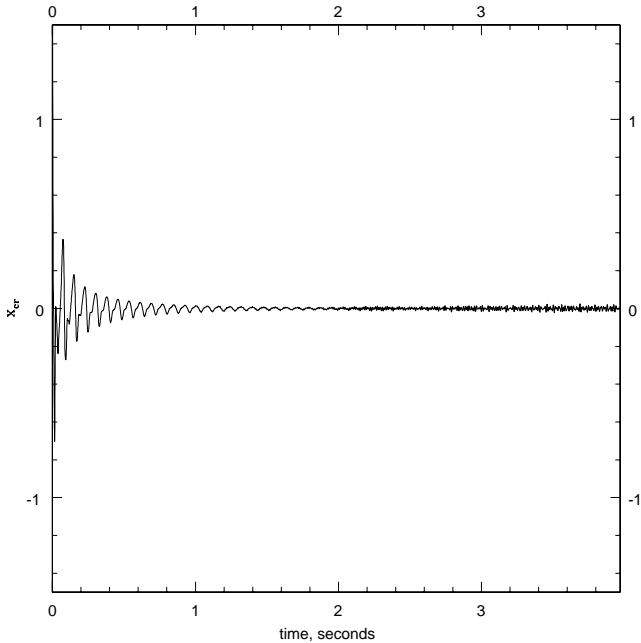, width=9cm}
%}
\end{center}
\caption{Crust displacement as a function of time.}
\end{figure}
Like in our toy models, we observe an initially rapid decay of crustal motion,
due to pumping of the crustal energy into the core. The crustal motion is then stabilized as the crust reacts 
to the vigorous movement of the neutron-star core. In Fig.~10
we plot the dynamical spectrum of the crustal motion for
the first 100 seconds (we split the time-axis into 1/2-second intervals, and take a Fourier transform at each interval.
The density of points represents the magnitude of the square of the Fourier transform). In the dynamical spectrum,
we see several QPO-type features. The low-frequency QPOs asymptote to the turning points of the core continuum,
$c_a n\pi/R$ (here we are considering the odd modes). Intermittent drifting QPOs appear at higher frequencies, and
they get strongly amplified near the crustal frequencies. The nature of the QPO frequency drifts is unclear to us at
this point.
\begin{figure}
\begin{center}
%\hbox{
\epsfig{file=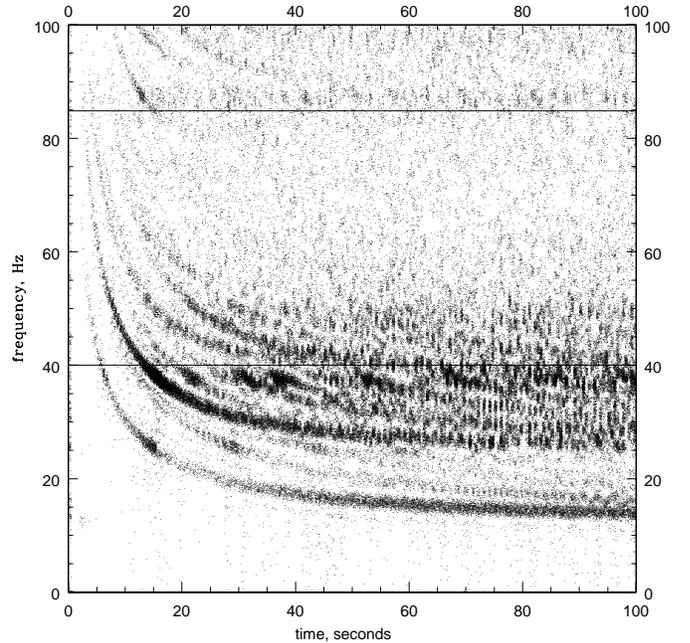, width=9cm}
%}
\end{center}
\caption{Dynamical spectrum of the crustal displacement. Thin horizontal lines mark the pure crustal
frequencies. Two low-frequecy QPOs asymptote to the continuum turning points, and are unrelated
to the crustal modes. 
Larger version of the figure, with better resolution near the crustal frequencies, is
available upon request.}
\end{figure}

The simulated dynamical spectrum of Fig.~10 is in qualitative agreement with that of the x-ray lightcurve in
the tail of a giant flare, see Israel et al.~2005. In both cases there is significant steady power below the lowest crustal
frequency, and we conclude that the observed $18$Hz QPO is almost certainly the turning point of the core continuum.
We also see an intermittent excess of power near the crustal frequencies, in qualitative agreement with the observations.

\section{Outlook}
In this paper we have elucidated the crucial role that
the core Alfven continuum plays in the dynamics of torsional motion of magnetars.
We have shown that the global torsional modes do not exist unless the friction is unphysically
large. We have performed a series of initial-value simulations for a simple but geometrically
realistic magnetar model, and have observed QPOs with the properties closely resembling
those in the tails of giant magnetar flares. In our model, the steady low-frequency QPOs are associated with
the turning points of the Alfven continuum.
This gives us a constraint on the combination of magnetic field geometry and strength and density of the core material
which is coupled to the magnetic field. For our geometry,
\begin{equation}
\left({B\over 10^{15}\hbox{G}}\right)
\left({10^{14}\hbox{g}/\hbox{cm}^3\over \rho}\right)^{1/2}
\left({10\hbox{km}\over R}\right)\simeq 1.
\label{constraint}
\end{equation}
Some of the higher-frequency power is clustered around the crustal frequencies, in agreement with the
observations.  This seems to be due
to intermittent amplification of the drifting QPOs when they are close to a crustal frequency. 

While the qualitative agreement between our simulations and observations is good, we see several directions
for future research: \newline 1.~Investigate qualitatively the origin of the QPO drifts is the simulations,
and the reason why the QPO amplitudes get amplified near the crustal frequencies.
\newline 2.~Study the Alfven continuum for more realistic field geometries, e.g.~the ones proposed in
Braithwaite \& Spruit 2004, 2006.
\newline 3.~Investigate quantitatively the effect of viscous 
friction
in the core. We have done some pilot studies for the toy models in the subsection 3.1, and have found that
viscous friction is very efficient in draining of the core's kinetic energy, but does not significantly
affect the QPOs. It is interesting to learn whether some of the long-term thermal afterglow could
be generated from the heat deposited in the core by viscously damped Alfven waves.
\newline 4.~Investigate the dynamics of the magnetosphere. It is likely that the magnetosphere also 
features the continuum of Alfven modes, and they will affect the emission of x-rays in the giant flare
afterglow.

\section{acknowledgments}
I thank  Wim van Saarlos and Henk Spruit for helpful 
discussions. The final stages of this work were completed at the 
Australia Telescope National Facility,
Sydney, for whose hospitality I am grateful.

\section{Appendix A: elastic forces and accelerations}
The purpose of this Appendix is to derive the expression (\ref{L}) in the text for
the crustal acceleration due to elastic forces. Since our problem is axially symmetric,
the quickest derivation is obtained by considering the flow of the $z$-component of
the shell's angular momentum density.
The displacement $\bar{\xi}_\phi(\theta)$ causes the horizontal 
shear stress in the shell which is given by:
\begin{equation}
T_{\hat{\theta}\hat{\phi}}=(\mu/R) \sin\theta {\partial(\bar{\xi}_\phi(\theta)/\sin\theta)
      \over\partial\theta},
\label{T1}
\end{equation}
where $\mu$ is the shear modulus of the shell. The flow of angular momentum in the $\partial_\theta$
direction is given by 
\begin{equation}
\hbox{Lflux}=
2\pi (R \sin\theta)^2 \int dr T_{\hat{\theta}\hat{\phi}},
\label{flow}
\end{equation}
where the integration is
over the thickness of the shell. The small thickness of the shell warrants the assumption
of $r$-independent  $\bar{\xi}_\phi(\theta)$, so only $\mu$ needs to be integrated. The angular momentum 
density with respect to the $R\theta$ coordinate is given by 
\begin{equation}
\hbox{Ldensity}=2\pi (R\sin\theta)^2\Sigma 
(\partial \bar{\xi}_\phi(\theta)/\partial t).
\label{Ldensity}
\end{equation}
Angular momentum conservation demands
\begin{equation}
\partial (\hbox{Ldensity})/\partial t=-(1/R)\partial (\hbox{Lflux})/\partial\theta.
\label{lcons}
\end{equation}
By substituting Eqs (\ref{flow}) and (\ref{Ldensity}) into Eq.~(\ref{lcons}) and performing
some trivial algebra, one obtains
\begin{equation}
{\partial^2\bar{\xi}_\phi(\theta)\over \partial t^2}=\omega_{\rm el}^2
\left[{\partial^2\bar{\xi}_\phi\over\partial \theta^2}+
\cot(\theta){\partial\bar{\xi}_\phi\over\partial\theta}-(\cot^2(\theta)-1)\bar{\xi}_\phi
\right],
\label{elas}
\end{equation}
where 
\begin{equation}
\omega_{\rm el}^2={\int \mu dr\over \Sigma R^2}={\bar{\mu}/\bar{\rho}\over R^2}.
\label{omel}
\end{equation}
The right-hand side of Eq.~(\ref{elas}) is the operator $L(\bar{\xi}_{\phi})$ used in the text.

\section{Appendix B: QPOs from edges and turning points in the continuum}
The purpose of this Appendix is to explain  mathematically the late-time behaviour of QPOs produced
by the edges and turning points of the continuum in the toy models of subsection 3.1.
\subsection{Edges}
 Lets assume that after initial excitation, 
 the oscillation amplitude of the small pendulae described in the subsection 3.1
 does not change. Then the  force acting on the
 big pendulum can be written as
 \begin{equation}
 F(t)=\int_{\omega_{\rm min}}^{\omega_{\rm max}} d\omega f(\omega)\exp(i\omega t),
 \label{F}
 \end{equation}
where $\omega_{\rm min}$ and $\omega_{\rm max}$ are the upper and lower edges of the continuum,
and function $f(\omega)$ encompasses the amplitude of excitation, the coupling strength, and the 
density of the continuum states at frequency $\omega$. Lets continue
$f(\omega)$ smoothly outside the $(\omega_{\rm min}, \omega_{\rm max})$ region in such a way that
$f\rightarrow 0$ sufficiently fast (to be presently specified) as $|\omega|\rightarrow\infty$. We
call this new function $\tilde{f}(\omega)$ and choose it in such a way that
its inverse Fourier transform, $f^*(t)$, has a finite
effective width $\Delta t$. Then $F(t)$ is the convolution of $f^*(t)$ and $s^*(t)$,
\begin{equation}
F(t)\propto \int d\tau f^*(\tau) s^*(t-\tau).
\label{F1}
\end{equation}
Here 
\begin{equation}
s^*(t)=[\exp(i\omega_{\rm max}t)-\exp(i\omega_{\rm min}t)]/t
\label{s*}
\end{equation}
is the inverse Fourier transform of the function $s(\omega)$, which equals $1$ for
$\omega_{\rm min}<\omega<\omega_{\rm max}$,
and $0$ otherwise.
When $t\gg\Delta t$, we can substitute $1/(t-\tau)$ with $1/t$ in Eq.~(\ref{F1}).
Then we have for late times
\begin{equation}
F(t)\propto [f(\omega_{\rm max})\exp(i\omega_{\rm max}t)-
             f(\omega_{\rm min})\exp(i\omega_{\rm min}t)]/t,
\label{F2}
\end{equation}
and thus the edge QPOs decay as $1/t$.

\subsection{Turning points}
Consider a downward turning point at angular frequency $\omega_0$ (i.e., a local
maximum in $\omega$ as a function of $m$). The density of states near the turning point 
scales as $(\omega_0-\omega)^{-1/2}$. Then the contribution of the oscillators near the turning
point
frequency to the force acting on the big pendulum is given by
\begin{equation}
F(t)\propto \int^{\omega_0}d\omega (\omega_0-\omega)^{-1/2}\exp(i\omega t).
\label{F3}\end{equation}
Introducing the new variable $x=(\omega_0-\omega)t$ and noticing that for large $t$ the range of
$x$ becomes essentially $(-\infty, 0)$, we see that
\begin{equation}
F(t)\propto \left[\int_{-\infty}^0 x^{-1/2}\exp(ix)\right]\times {\exp(i\omega_0 t)\over t^{1/2}}.
\label{F4}
\end{equation}
Thus, the amplitude of the QPO associated with the turning point decays as $t^{-1/2}$.

\end{document}